\documentclass[aps,prb,reprint,notitlepage,amsmath,amssymb]{revtex4-2}
\usepackage{graphicx,bm,booktabs}
\usepackage[colorlinks=true,linkcolor=blue,citecolor=blue]{hyperref}
\begin{document}
\title{Exact magnetic ground states and competing interlayer exchanges in layered kagome magnets}
\author{Vladimir Antropov}
\affiliation{Ames National Laboratory, Ames, Iowa 50010, USA}
\begin{abstract}
Magnetic ground states in layered kagome magnets are investigated using a Heisenberg model with six interlayer exchange interactions. When the two magnetic sublattices are equivalent, in the sense that their diagonal Fourier exchange functions coincide, the Luttinger--Tisza method gives the exact classical ground state for arbitrary exchange range: a coplanar single-wave-vector spiral. Extending the three-exchange model of Rosenfeld and Mushnikov shows that longer-range interactions can change the preferred magnetic order from spiral to ferromagnetic and back to spiral, rather than merely modifying the spiral period. We calculate phase diagrams, harmonic stability, magnetic-field response, and the sensitivity of the ordering wave vector to individual exchanges. The propagation vector, internal layer angle, saturation field, and planar susceptibility constrain at most four combinations of the six exchange constants, and one of these only weakly. The results connect first-principles interlayer exchanges with the selection of noncollinear magnetic order.
\end{abstract}
\maketitle
\section{Introduction}\label{sec:intro}
Structural stability and a sizable calculated magnetic moment do not by themselves determine the magnetic ground state of a material. This distinction is particularly important for layered kagome compounds, where magnetic order affects the symmetry of electronic states and can be changed by composition or applied field. Recent reviews describe the diversity of magnetic configurations and their relationship to topological and transport properties in the kagome families~\cite{XuYinQuJia2023,Negi2025,Sharma2026}. Experiments on $RMn_6Sn_6$ materials establish that competition between interlayer interactions can produce incommensurate order and complex field-induced structures~\cite{Venturini1996,Ghimire2020,Wang2021,Dally2021,ErMn2024,HighField2025}. The magnetic stacking along the crystallographic $c$ axis is therefore central to understanding these compounds.

The practical need has grown with computational materials discovery. High-throughput density-functional calculations have examined 312 substitutional $AT_6X_6$ compositions and identified additional stable magnetic candidates~\cite{Chen2026}. Searches of related $AT_6X_4$ and $AT_6X_5$ structure types broaden the range of candidate compounds~\cite{ChenArxiv2026}. Such studies typically examine a limited number of candidate magnetic structures, frequently including collinear arrangements. A lowest-energy collinear solution, however, need not be the lowest-energy state of a frustrated interlayer exchange model. Competing exchanges and nearly degenerate collinear states make noncollinear order a plausible alternative, although not every predicted compound is expected to be a spiral magnet.

The three-exchange double-layer model of Rosenfeld and Mushnikov (RM) provides a reference for $RMn_6X_6$ magnets~\cite{RosenfeldMushnikov2008}. It describes the competition between inequivalent adjacent-layer couplings and a further interlayer interaction; subsequent work examined nested spirals and their stability~\cite{Rosenfeld2009Nested,Rosenfeld2009Spiral}. First-principles calculations can now resolve more distant exchange paths, whose effects need not reduce to a change of spiral period. We therefore retain six interlayer exchanges and examine how they determine the magnetic stacking, stability, and field response. Both the turn angle between repeats and the relative orientation of the two layers are allowed to relax.

We use the Luttinger--Tisza (LT) method~\cite{LuttingerTisza1946,LyonsKaplan1960,KaplanMenyuk2007}. For the isotropic classical model, the equivalent Mn planes of $RMn_6Sn_6$ have identical diagonal Fourier exchange functions, so the LT eigenmode satisfies the fixed-length constraint. This gives the exact zero-field ground state within the rigid-layer model. The exact exchange spectrum determines the phase diagrams and provides the starting point for the stability and magnetic-response calculations. The model assumes equal-length layer moments and omits rare-earth magnetism, longitudinal fluctuations, and additional anisotropic or non-Heisenberg interactions.

\section{Six-exchange two-sublattice Heisenberg model}\label{sec:model}
\subsection{Real-space exchange geometry and conventions}
\begin{figure*}[t]
\centering
\includegraphics[width=0.66\textwidth]{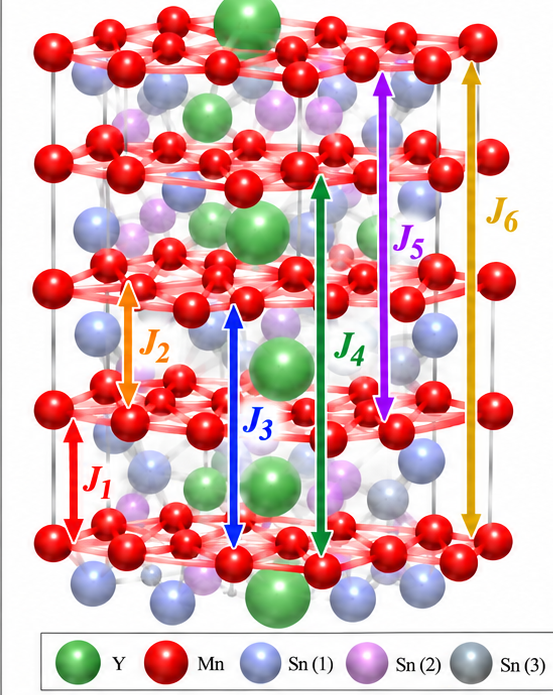}
\caption{Six effective interlayer exchanges in the two-Mn-plane kagome model, illustrated on a YMn$_6$Sn$_6$-inspired layer stack. The arrows indicate the six effective interlayer exchange paths, not individual microscopic Mn--Mn bonds. The same-sublattice terms $J_3$ and $J_6$ also contain the corresponding $B$--$B$ interactions, omitted for clarity. The nonmagnetic atoms are illustrative.}\label{fig:geometry}
\end{figure*}

The model consists of two magnetic layers, $A$ and $B$, in each repeat $n$. The unit vectors $\bm e_{An}$ and $\bm e_{Bn}$ give their spin directions. Counting each exchange bond once, the energy is
\begin{align}\label{eq:H}
\mathcal H={}&-\sum_n\bigl[J_1\bm e_{An}\!\cdot\!\bm e_{Bn}
+J_2\bm e_{Bn}\!\cdot\!\bm e_{A,n+1}\nonumber\\
&+J_4\bm e_{An}\!\cdot\!\bm e_{B,n+1}
+J_5\bm e_{Bn}\!\cdot\!\bm e_{A,n+2}\nonumber\\
&+J_3(\bm e_{An}\!\cdot\!\bm e_{A,n+1}+\bm e_{Bn}\!\cdot\!\bm e_{B,n+1})\nonumber\\
&+J_6(\bm e_{An}\!\cdot\!\bm e_{A,n+2}+\bm e_{Bn}\!\cdot\!\bm e_{B,n+2})\bigr].
\end{align}
The first three bonds reproduce the established three-exchange layered model~\cite{RosenfeldMushnikov2008}. The $J_4$, $J_5$, and $J_6$ paths describe additional interlayer interactions. Positive $J_a$ favors parallel alignment on the corresponding bond. The exchange labels denote the particular paths shown in Fig.~\ref{fig:geometry}, not a universal shell ordering. All energies below are per two-layer crystallographic repeat.
\subsection{Exact homogeneous spiral energy}
A planar spiral with a constant rotation angle between successive two-layer repeats is described by
\begin{align}\label{eq:ansatz}
\bm e_{An}&=(\cos n\Phi,\sin n\Phi,0),\nonumber\\
\bm e_{Bn}&=(\cos(n\Phi+\delta),\sin(n\Phi+\delta),0).
\end{align}
Here $\Phi=2\pi q$ is the rotation between consecutive crystallographic repeats and $\delta$ is the internal $A$--$B$ phase. Direct substitution into Eq.~\eqref{eq:H} yields
\begin{align}\label{eq:Ephi_delta}
E(\Phi,\delta)={}&-J_1\cos\delta-J_2\cos(\Phi-\delta)
-J_4\cos(\Phi+\delta)\nonumber\\
&-J_5\cos(2\Phi-\delta)-2J_3\cos\Phi-2J_6\cos2\Phi.
\end{align}
The four bonds connecting $A$ and $B$ layers can be collected into $C(\Phi)$, while $D(\Phi)$ contains the bonds between like layers:
\begin{align}\label{eq:C}
C(\Phi)&=J_1+J_2e^{-i\Phi}+J_4e^{i\Phi}+J_5e^{-2i\Phi},\\
D(\Phi)&=2J_3\cos\Phi+2J_6\cos2\Phi.\label{eq:D}
\end{align}
The energy is $E=-D-\mathrm{Re}[e^{i\delta}C]$. If $C\ne0$, the minimizing phase satisfies
\begin{equation}\label{eq:delta}
e^{i\delta_*(\Phi)}=\frac{C(\Phi)^*}{|C(\Phi)|},
\end{equation}
and hence the lowest homogeneous-spiral energy is
\begin{equation}\label{eq:Eopt}
\boxed{E_0(\Phi)=-2J_3\cos\Phi-2J_6\cos2\Phi-|C(\Phi)|.}
\end{equation}
If $C=0$, the intersublattice exchange does not determine $\delta$. The dependence of the energy on the spiral wave vector may then develop a cusp, where the derivative formulas cannot be used.
\subsection{Momentum-space exchange matrix}
With the Fourier convention consistent with Eq.~\eqref{eq:C}, the Hermitian exchange matrix is
\begin{equation}\label{eq:Jmatrix}
\mathsf J(\Phi)=\begin{pmatrix}D(\Phi)&C(\Phi)\\C(\Phi)^*&D(\Phi)\end{pmatrix}.
\end{equation}
Its eigenvalues are
\begin{equation}\label{eq:eigenvalues}
\lambda_\pm(\Phi)=D(\Phi)\pm|C(\Phi)|.
\end{equation}
The upper exchange eigenvalue reproduces the optimized spiral energy, $E_0=-\lambda_+$. Its relation to the unrestricted classical ground state follows from the LT argument in Sec.~\ref{sec:arbitrary}.

\section{Exact ground states for arbitrary-range exchange networks}\label{sec:arbitrary}
The six exchanges in Eq.~\eqref{eq:H} describe the selected interlayer bonds. The ground-state result below also applies when more distant bonds are included. Consider an isotropic classical Heisenberg model with two unit spins per primitive cell and translationally invariant bilinear exchanges $J_{\alpha\beta}(\bm R)$, where $\alpha,\beta=A,B$ and $\bm R$ connects primitive cells. Each physical bond is counted once, with $J_{\beta\alpha}(-\bm R)=J_{\alpha\beta}(\bm R)$. The exchange matrix is
\begin{equation}\label{eq:generalJ}
\mathsf J(\bm q)=\begin{pmatrix}
D_A(\bm q)&C(\bm q)\\ C(\bm q)^*&D_B(\bm q)
\end{pmatrix},\quad
\begin{aligned}
D_\alpha(\bm q)&=\sum_{\bm R}J_{\alpha\alpha}(\bm R)e^{i\bm q\cdot\bm R},\\
C(\bm q)&=\sum_{\bm R}J_{AB}(\bm R)e^{i\bm q\cdot\bm R}.
\end{aligned}
\end{equation}
Here the exchange-matrix convention includes both directions of each like-sublattice bond; the Hamiltonian carries the corresponding factor $1/2$. Sublattice positions may instead be included in the Fourier phases by a change of basis, without changing the eigenvalues. For \emph{equivalent} sublattices, in the precise sense that $D_A(\bm q)=D_B(\bm q)\equiv D(\bm q)$ for every $\bm q$, the upper eigenvalue and an associated normalized eigenvector are
\begin{align}\label{eq:general2}
\lambda_+(\bm q)&=D(\bm q)+|C(\bm q)|,\nonumber\\
\bm v_+(\bm q)&=\frac{1}{\sqrt2}
\begin{pmatrix}1\\ C(\bm q)^*/|C(\bm q)|\end{pmatrix}.
\end{align}
The expression for the eigenvector assumes $C\ne0$; at $C=0$ the two sublattice phases are degenerate. In particular, equality of the diagonal functions does \emph{not} require equality of the different $A$--$B$ bonds.

\emph{Exact zero-field ground state.} The following statement is an application of the LT method~\cite{LuttingerTisza1946,LyonsKaplan1960,KaplanMenyuk2007}. The Fourier bound below is the LT ``weak constraint'', and the equal sublattice amplitudes of the optimal eigenvector ensure that the ``strong constraint'' of unit spin length is satisfied. Single-wave-vector helical ground states of Bravais-lattice Heisenberg models were obtained in the same way by Yoshimori~\cite{Yoshimori1959}. We state the two-sublattice form explicitly because it is used throughout. Consider the isotropic bilinear Hamiltonian
\begin{equation}\label{eq:theoremH}
\mathcal H=-\frac12\sum_{\bm R,\bm R'}\sum_{\alpha,\beta=A,B}
J_{\alpha\beta}(\bm R-\bm R')\,
\bm e_{\bm R\alpha}\cdot\bm e_{\bm R'\beta}.
\end{equation}
Every spin satisfies $|\bm e_{\bm R\alpha}|=1$. Assume translational invariance, a well-defined exchange Fourier transform, and $D_A(\bm q)=D_B(\bm q)$ at every wave vector. Then at least one exact global zero-field ground state is a coplanar single-$\bm Q$ spiral, with
\begin{equation}\label{eq:generalground}
\boxed{\frac{E_{\rm GS}}{N_c}=-\max_{\bm q}\bigl[D(\bm q)+|C(\bm q)|\bigr].}
\end{equation}
Here $N_c$ is the number of primitive cells, and $\bm Q$ maximizes the expression in brackets.

\emph{Proof.} We use the Fourier convention
$\bm e_{\bm q\alpha}=N_c^{-1/2}\sum_{\bm R}
\bm e_{\bm R\alpha}e^{-i\bm q\cdot\bm R}$.
The largest exchange eigenvalue limits the exchange energy of every Fourier component:
$\bm e_{\bm q}^{\dagger}\mathsf J(\bm q)\bm e_{\bm q}
\leq\Lambda\bm e_{\bm q}^{\dagger}\bm e_{\bm q}$,
where $\Lambda=\max_{\bm q}\lambda_+(\bm q)$. Summing over wave vectors and using the fixed spin lengths gives
\begin{align}\label{eq:global_bound}
\mathcal H&=-\frac12\sum_{\bm q}
\bm e_{\bm q}^{\dagger}\mathsf J(\bm q)\bm e_{\bm q}
\geq-\frac{\Lambda}{2}\sum_{\bm q}
\bm e_{\bm q}^{\dagger}\bm e_{\bm q}\nonumber\\
&=-N_c\Lambda.
\end{align}
No spin configuration can have an exchange energy below this value, including nonuniform, noncoplanar, and multi-$\bm q$ states. A planar spiral reaches this energy. At the preferred wave vector $\bm Q$, we take
$e^{i\delta}=C(\bm Q)^*/|C(\bm Q)|$ and write
\begin{align}\label{eq:generalspiral}
\bm e_{A\bm R}&=(\cos\bm Q\!\cdot\!\bm R,
\sin\bm Q\!\cdot\!\bm R,0),\nonumber\\
\bm e_{B\bm R}&=(\cos(\bm Q\!\cdot\!\bm R+\delta),
\sin(\bm Q\!\cdot\!\bm R+\delta),0).
\end{align}
Each spin has unit length. The sublattice phase vector
$(1,e^{i\delta})^T$ is an eigenvector of
$\mathsf J(\bm Q)$ with eigenvalue $\Lambda$, and its complex conjugate gives the corresponding mode at $-\bm Q$. The energy of this planar spiral therefore equals the lowest energy allowed by Eq.~\eqref{eq:global_bound}, establishing Eq.~\eqref{eq:generalground}. When $C(\bm Q)=0$, either relative phase is allowed and the same conclusion follows. \emph{Q.E.D.}

The planar spiral is an exact ground state, although other magnetic structures may have the same energy: distinct maximizing wave vectors may yield additional degenerate states. Finite periodic samples require a compatible $\bm Q$; the unrestricted continuous-wave-vector statement holds in the thermodynamic limit. Arbitrarily distant $AA$, $BB$, and $AB$ exchanges only add Fourier harmonics to $D$ and $C$, without enlarging the $2\times2$ eigenvalue problem. Longer exchange range can nevertheless generate additional competing ordering vectors.

\emph{General number of sublattices.} For $s$ unit spins per cell, let $\lambda_{\max}(\bm q)$ denote the largest eigenvalue of the $s\times s$ Hermitian exchange matrix. The same argument gives
\begin{equation}\label{eq:generalNbound}
\frac{E_{\rm GS}}{N_c}\ge-\frac{s}{2}\max_{\bm q}\lambda_{\max}(\bm q).
\end{equation}
A single-$\bm Q$ spiral reaches the lowest possible exchange energy if the exchange eigenvector at the preferred wave vector has $|v_\alpha|=1/\sqrt s$ for \emph{every} sublattice, which is the generalized LT condition~\cite{LyonsKaplan1960,KaplanMenyuk2007}. The relative phases of its components then describe a coplanar spiral with unit spins on every sublattice and the lowest possible exchange energy. Other ground states are also possible: combinations of degenerate eigenmodes may also satisfy the local constraints. For two inequivalent sublattices with $D_A\ne D_B$, the maximizing eigenvector generally has unequal amplitudes, so the Fourier maximum alone need not give the ground state.

The result is exact for the stated classical Heisenberg model, but does not establish the validity of a rigid-layer reduction for a material. Additional magnetic sites within a kagome plane, unequal moment lengths, non-Heisenberg interactions, anisotropy, applied fields, and longitudinal fluctuations require separate treatment. In particular, a two-layer ground state can still be unstable to intralayer degrees of freedom omitted when each plane is represented by one spin.

\section{Ground-state determination and exchange dependence}\label{sec:general}
\subsection{A useful form of the spiral energy}
Writing $x=\cos\Phi$, the intersublattice exchange contribution takes the form
\begin{align}\label{eq:P}
P_3(x)&=|C(\Phi)|^2=a_0+a_1x+a_2x^2+a_3x^3,\\
a_0&=J_1^2+J_2^2+J_4^2+J_5^2-2(J_1J_5+J_2J_4),\nonumber\\
a_1&=2[J_1(J_2+J_4)+J_2J_5]-6J_4J_5,\nonumber\\
a_2&=4(J_1J_5+J_2J_4),\nonumber\\
a_3&=8J_4J_5.\nonumber
\end{align}
Thus, for $P_3(x)>0$,
\begin{equation}\label{eq:Ex}
E_0(x)=-2J_3x-2J_6(2x^2-1)-\sqrt{P_3(x)}.
\end{equation}
This form is convenient for determining the preferred spiral period.
\subsection{Finding the preferred spiral period}
In the interior interval $0<\Phi<\pi$, homogeneous spirals satisfy $dE_0/d\Phi=0$. Since $dx/d\Phi=-\sin\Phi\ne0$, this is equivalent to
\begin{equation}\label{eq:stationarity}
-2J_3-8J_6x-\frac{P_3'(x)}{2\sqrt{P_3(x)}}=0.
\end{equation}
The possible equilibrium spiral periods can also be found from the polynomial equation
\begin{equation}\label{eq:stationarityquintic}
4(2J_3+8J_6x)^2P_3(x)-[P_3'(x)]^2=0.
\end{equation}
The equation is generally fifth order in $x$. Squaring the equilibrium condition introduces additional roots that need not satisfy Eq.~\eqref{eq:stationarity}. Only solutions with $-1<x<1$ and $P_3(x)>0$ that satisfy the original equilibrium condition are retained, Eq.~\eqref{eq:stationarity}. Their energies are compared with those of the collinear states $\Phi=0$ and $\Phi=\pi$. At $P_3(x)=0$, direct energy comparison is necessary because the derivative may be discontinuous.
Introducing $Q(x)=[P_3'(x)]^2-2P_3(x)P_3''(x)$, the second derivative of the spiral energy is
\begin{equation}\label{eq:curvature}
\frac{d^2E_0}{dx^2}=-8J_6+\frac{Q(x)}{4P_3(x)^{3/2}}.
\end{equation}
At an equilibrium spiral wave vector, $d^2E_0/d\Phi^2=\sin^2\Phi\,d^2E_0/dx^2$. A positive second derivative indicates that the spiral resists a small change in its period. The sign of this curvature distinguishes a local pitch minimum from a maximum.
\subsection{Ground-state selection}
The ground state has the lowest energy among the stationary spirals, collinear endpoints, and cusp configurations $P_3(x)=0$. Its internal layer angle follows from $C(\Phi)$ when $C\ne0$.

\section{Ground-state phase diagrams}\label{sec:phase}
\subsection{Exact three-exchange reference and phase nomenclature}
The RM phase diagrams provide a reference~\cite{RosenfeldMushnikov2008}. Their Fig.~2 shows separate results for $J_1>0$ and $J_1<0$. The same axes $x=J_2/J_1$ and $y=J_3/J_1$ are retained, with $-3\leq x\leq3$ and $-2\leq y\leq2$, and their labels F, AF1, AF2, AF3, H1, and H2. The first four denote distinct collinear stacking patterns; H1 and H2 denote spiral structures for which the two consecutive turn angles have the same and opposite signs, respectively. In particular, H1 and H2 do \emph{not} denote two distinct minima of the spiral energy at fixed exchange parameters. In RM's Table~1 the collinear assignments are F: $(\delta,\Phi)=(0,0)$, AF1: $(\pi,0)$, AF2: $(0,\pi)$, and AF3: $(\pi,\pi)$. These assignments are used throughout. In the corresponding layer sequences, AF1 alternates within each repeat, whereas AF2 alternates between repeats. 

With $J_4=J_5=J_6=0$, the present Hamiltonian reduces exactly to their zero-field exchange energy,
\begin{equation}\label{eq:rosenfeld_energy}
 E_{\rm RM}(\Phi,\delta)=-J_1\cos\delta-J_2\cos(\Phi-\delta)-2J_3\cos\Phi.
\end{equation}
The optimized energy is $E_{\rm RM,0}(\Phi)=-2J_3\cos\Phi-|J_1+J_2e^{-i\Phi}|$. For an interior stationary pitch, writing $u=\cos\Phi$, the stationary equation implies
\begin{equation}\label{eq:rosenfeld_pitch}
 u=\frac{J_1^2J_2^2/(4J_3^2)-J_1^2-J_2^2}{2J_1J_2},\qquad -1<u<1,
\end{equation}
provided $J_1J_2J_3\neq0$. The unsquared stationary equation additionally requires $J_1J_2/J_3<0$; a value of $u$ in $(-1,1)$ alone is not sufficient. The spiral solution is compared with the two collinear endpoints, $\Phi=0$ and $\pi$, and the lowest energy is selected. The continuous spiral--collinear boundaries, in RM notation, lie on four curves
\begin{align}\label{eq:rosenfeld_boundaries}
 k:\ &x+2y(x+1)=0, &l:\ &x-2y(x+1)=0,\\
 m:\ &x-2y(x-1)=0, &n:\ &x+2y(x-1)=0.\nonumber
\end{align}
Only the parts of these curves separating actual lowest-energy states are phase boundaries. Their assignments to adjacent phases differ between the two signs of $J_1$, as specified in Table~2 of Ref.~\cite{RosenfeldMushnikov2008}.

\subsection{Representative phase diagrams}
\begin{figure*}[t]
\centering\includegraphics[width=0.94\textwidth]{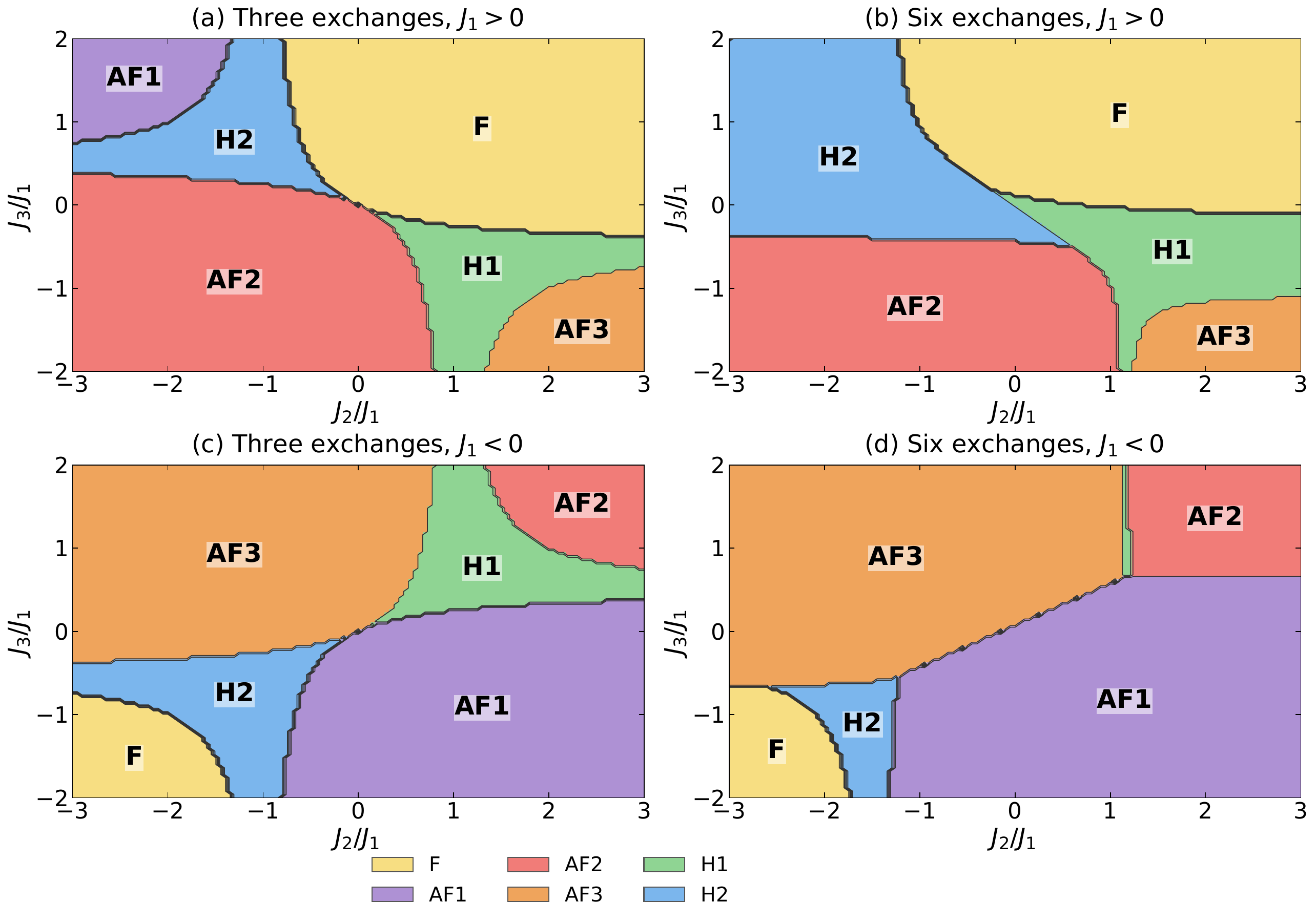}
\caption{Ground-state phase diagrams in the RM coordinates~\cite{RosenfeldMushnikov2008}. Left column: numerically evaluated three-exchange model, $J_4=J_5=J_6=0$; right column: calculated six-exchange sections at $J_4/J_1=0.15$, $J_5/J_1=0.30$, and $J_6/J_1=-0.18$. Upper row: $J_1>0$; lower row: $J_1<0$. F and AF1--AF3 denote the four collinear stacking patterns, while H1 and H2 distinguish the sign relation of consecutive spiral turn angles, as in the original paper. Phase boundaries are resolved on a finite numerical grid.}\label{fig:rosenfeldcompare}
\end{figure*}

Figure~\ref{fig:rosenfeldcompare} places the original three-exchange diagram beside a six-exchange example. Both use the same axes and phase labels. The upper and lower rows correspond to $J_1>0$ and $J_1<0$, respectively. For the extended panels, the parameters are fixed at $J_4/J_1=0.15$, $J_5/J_1=0.30$, and $J_6/J_1=-0.18$, and $x$ and $y$ are varied. These values illustrate the model; they are not fitted to experiment. For the three-exchange panels, the analytical equilibrium-pitch expression was checked by directly minimizing the exchange energy. For the six-exchange panels, the exchange energy was minimized numerically with respect to the spiral period. For the extended model, the upper exchange eigenvalue is compared at the two collinear wave vectors and at the interior maxima. Phase boundaries separate states with equal ground-state energies within the numerical resolution.
The additional exchanges reshape the collinear and spiral regions. The phase labels identify the globally minimizing single-$q$ states, as guaranteed by the equivalent-layer Luttinger--Tisza construction. The dark lines mark changes in the minimizing state. Boundaries between collinear states are straight lines that follow from the endpoint eigenvalues $\lambda_+(0)=2J_3+2J_6+|C(0)|$ and $\lambda_+(\pi)=-2J_3+2J_6+|C(\pi)|$. In panel (d), the collinear equal-energy conditions give $C(\pi)=0$, or $J_2/J_1=1.15$, for AF2 and AF3, and $J_3/J_1=[|C(\pi)|-|C(0)|]/4J_1=0.65$ for AF1 and AF2 in the relevant parameter sector. These straight lines represent global phase boundaries only along segments where no spiral has lower energy. In particular, the narrow H1 region near $J_2/J_1=1.15$ interrupts the simple collinear assignment: the optimized spiral energy is lower there.

\subsection{Distinct roles of the longer-range exchanges}\label{sec:atlas}
The three added interactions modify the spiral in physically different ways. The
$J_4$ path contributes $-J_4\cos(\Phi+\delta)$ and changes the complex
intersublattice exchange amplitude; $J_5$ contributes
$-J_5\cos(2\Phi-\delta)$ and adds a longer-range intersublattice harmonic.
By contrast, the equivalent-layer $J_6$ coupling contributes
$-2J_6\cos2\Phi$ to the diagonal exchange spectrum. Thus $J_6$ changes
the optimal pitch without directly changing the minimizing internal angle
at a \emph{fixed} pitch. The resulting pitch and internal angle must therefore be determined together. Separate sweeps of each
coupling for both signs of $J_1$, including the numerical phase-atlas
conventions and limitations, are given in the Supplemental Material~\cite{SM}.
They demonstrate changes of the collinear--spiral regions, but a change
of phase label alone does not establish the number of competing local
spiral minima.

\section{Controlled progression from three to six exchanges}\label{sec:progression}
Their combined effect is examined by holding $(J_1,J_2,J_3)/J_1=(1,0.25,-0.12)$ fixed and successively adding $J_4/J_1=0.025$, $J_5/J_1=0.045$, and $J_6/J_1=-0.035$. Each step adds one nonzero interaction without readjusting the preceding ones. The exchange constants are illustrative, not fitted to a particular compound.

For every parameter set, the exact function $E_0(\Phi)$ in Eq.~\eqref{eq:Eopt} is minimized over $0\leq\Phi\leq\pi$, including both endpoints. The internal angle follows from Eq.~\eqref{eq:delta}. Table~\ref{tab:progression} lists the resulting ground-state quantities. For a perpendicular field, the reported saturation scale is $B_{\rm sat}M_s=\Lambda-\lambda_+(0)+2K$, using the same $K/J_1=0.04$ for all four cases and a uniform ferromagnetic $k=0$ mode. This field scale is derived in Sec.~\ref{sec:rmfieldcomparison}.
\begin{table}[t]
\caption{Controlled three-to-six-exchange sequence, with $J_1=1$, $J_2=0.25$, $J_3=-0.12$, and successively added $(J_4,J_5,J_6)=(0.025,0.045,-0.035)$. Energy is per two-layer repeat. The saturation field uses $K=0.04$. Values are obtained by global minimization of Eq.~\eqref{eq:Eopt}.}\label{tab:progression}
\begin{ruledtabular}
\begin{tabular}{lrrrr}
Model & $q$ & $\delta$ (rad) & $E_0/J_1$ & $B_{\rm sat}M_s/J_1$\\
$J_1$--$J_3$ & 0.2428 & 0.2421 & $-1.0308$ & 0.1008\\
$J_1$--$J_4$ & 0.0990 & 0.1068 & $-1.0354$ & 0.0804\\
$J_1$--$J_5$ & 0 & 0 & $-1.0800$ & 0.0800\\
$J_1$--$J_6$ & 0.2033 & 0.2266 & $-1.0583$ & 0.1283
\end{tabular}
\end{ruledtabular}
\end{table}
The added $J_4$ reduces the spiral wave vector, while $J_5$ makes the ferromagnetic state energetically preferred. The negative $J_6$ then restores a nonzero spiral pitch. Table~\ref{tab:progression} also shows the accompanying changes in the internal angle and hard-axis saturation field. These are comparisons between distinct Hamiltonians, not phase transitions induced by continuously varying one parameter. The example demonstrates that longer-range exchange paths can change the magnetic order, not only the spiral period.

\section{Quantitative sensitivity to exchange interactions}\label{sec:quant_sensitivity}
The equilibrium pitch and internal phase respond together when an exchange parameter changes. For a spiral whose equilibrium angles vary continuously with exchange, define $\bm u=(\Phi,\delta)^T$ and the homogeneous-angle Hessian $\mathsf H_{ij}=\partial^2E/\partial u_i\partial u_j$. Differentiating the two equilibrium equations with respect to $J_a$ yields
\begin{equation}\label{eq:coupled_sensitivity}
\frac{\partial\bm u}{\partial J_a}=-\mathsf H^{-1}\frac{\partial}{\partial J_a}\begin{pmatrix}\partial_\Phi E\\\partial_\delta E\end{pmatrix}.
\end{equation}
For the four intersublattice bonds let $r=(0,-1,1,-2)$ correspond to $a=(1,2,4,5)$, and put $t_r=\delta+r\Phi$ and $w_r=J_r^{AB}\cos t_r$, where $J_r^{AB}=(J_1,J_2,J_4,J_5)$ in the stated order. The explicit Hessian is
\begin{equation}\label{eq:angle_hessian}
\mathsf H=\begin{pmatrix}
\sum_r r^2w_r+2J_3\cos\Phi+8J_6\cos2\Phi & \sum_r rw_r\\
\sum_r rw_r & \sum_r w_r
\end{pmatrix}.
\end{equation}
The mixed-derivative vector in Eq.~\eqref{eq:coupled_sensitivity} is $(r\sin t_r,\sin t_r)^T$ for $a=1,2,4,5$, $(2\sin\Phi,0)^T$ for $a=3$, and $(4\sin2\Phi,0)^T$ for $a=6$. These expressions determine the local changes of both angles directly from the equilibrium Hessian. The derivative of the reduced wave vector is $\partial q/\partial J_a=(2\pi)^{-1}\partial\Phi/\partial J_a$.

For a perpendicular field, when the uniform exchange eigenmode is ferromagnetic and $K$ is fixed, $h_{\rm sat}=B_{\rm sat}M_s=\lambda_+(\Phi_*)-\lambda_+(0)+2K$. Because the equilibrium pitch makes the first derivative of the energy vanish,
\begin{equation}\label{eq:field_sensitivity}
\frac{\partial h_{\rm sat}}{\partial J_a}=\left.\frac{\partial\lambda_+(\Phi)}{\partial J_a}\right|_{\Phi_*}-\frac{\partial\lambda_+(0)}{\partial J_a}.
\end{equation}
For the present parameter set $C(0)>0$, so these derivatives are $\cos t_r-1$ for $a=1,2,4,5$, $2(\cos\Phi-1)$ for $a=3$, and $2(\cos2\Phi-1)$ for $a=6$.

The derivatives are evaluated for the six-exchange endpoint of Table~\ref{tab:progression}. The parameters are $J_1=1$, $J_2=0.25$, $J_3=-0.12$, $J_4=0.025$, $J_5=0.045$, $J_6=-0.035$, and $K=0.04$. The minimum occurs at $\Phi=1.2772$ and $\delta=0.2266$, with $h_{\rm sat}=0.1283$. The eigenvalues of the homogeneous-angle Hessian are $0.1619$ and $1.0735$ in the chosen exchange-energy units, showing that small uniform changes of the two angles cost exchange energy.
For this parameter set, $J_6$ and $J_3$ have the largest local effects on the pitch, whereas $J_6$ has the largest effect on the saturation field. These derivatives describe a smooth spiral branch; they do not apply at a transition between competing minima. Numerical values and finite-difference checks are provided in the Supplemental Material.

\section{Harmonic stability of spiral states}\label{sec:stability}
\subsection{Changes of spin direction within the spiral plane}
A minimum of $E_0(\Phi)$ is stable against a uniform change of pitch, but the individual layers may also undergo nonuniform rotations. Writing
\begin{align}
\theta_{A,n}&=n\Phi+u_{A,n},\\
\theta_{B,n}&=n\Phi+\delta_*+u_{B,n},\qquad
\delta_*=-\arg C(\Phi).
\end{align}
The intersublattice bonds have relative displacements $r=0,-1,+1,-2$, respectively, and strengths $J_r^{AB}=(J_1,J_2,J_4,J_5)$. The corresponding exchange contributions to the in-plane spin-rotation energy are
\begin{equation}\label{eq:wr}
w_r=J_r^{AB}\cos(\delta_*+r\Phi),\qquad
W(p)=\sum_{r=0,-1,1,-2}w_r e^{irp}.
\end{equation}
For Fourier amplitudes $u_{\alpha,n}=N^{-1/2}\sum_p u_\alpha(p)e^{ipn}$, the energy through second order is
\begin{equation}\label{eq:quadratic_in}
\Delta E_{\parallel}=\frac12\sum_p
\bm u(p)^\dagger\mathsf H_{\parallel}(p)\bm u(p),
\end{equation}
with
\begin{align}\label{eq:Hparallel}
\mathsf H_{\parallel}(p)&=
\begin{pmatrix}A(p)&-W(p)\\-W(p)^*&A(p)\end{pmatrix},\\
A(p)&=\sum_r w_r+2J_3\cos\Phi(1-\cos p)\nonumber\\
&\quad+2J_6\cos2\Phi(1-\cos2p).
\label{eq:Aplanar}
\end{align}
For each wave vector $p$, the $A$ and $B$ layers can rotate mainly together or against one another. The corresponding coefficients of the quadratic spin-rotation energy are
\begin{equation}\label{eq:planar_eigs}
\lambda_{\parallel,\pm}(p)=A(p)\pm|W(p)|.
\end{equation}
At $p=0$, phase relaxation implies $\sum_r w_r=|C(\Phi)|\geq0$, so the acoustic branch obeys $\lambda_{\parallel,-}(0)=0$ and the optical branch has $\lambda_{\parallel,+}(0)=2|C(\Phi)|$. The zero-energy mode represents a common rotation of every spin in the plane. A spiral resists all small in-plane distortions only if $\lambda_{\parallel,-}(p)\geq0$ for \emph{every} $p$ in the Brillouin zone, not merely $E_0''(\Phi)>0$.

\subsection{Tilting spins out of the spiral plane}
Small out-of-plane rotations are described by
\begin{equation}\label{eq:zfluct}
\bm e_{\alpha n}=\sqrt{1-z_{\alpha n}^2}
(\cos\theta_{\alpha n},\sin\theta_{\alpha n},0)
+z_{\alpha n}\hat{\bm z},
\end{equation}
and retain terms quadratic in the deviations from the planar spiral. At an equilibrium spiral, the exchange field is parallel to each reference spin, with magnitude
\begin{equation}\label{eq:lambda0}
\Lambda(\Phi)=D(\Phi)+|C(\Phi)|=-E_0(\Phi).
\end{equation}
The exchange-energy change for out-of-plane spin rotations is described by
\begin{equation}\label{eq:Hperp}
\mathsf H_\perp(p)=\Lambda(\Phi)\mathsf I-\mathsf J(p),
\end{equation}
where $\mathsf J(p)$ is the exchange matrix in Eq.~\eqref{eq:Jmatrix}. Its eigenvalues are
\begin{align}\label{eq:perp_eigs}
\lambda_{\perp,-}(p)&=\Lambda(\Phi)-[D(p)+|C(p)|]\nonumber\\
&=E_0(p)-E_0(\Phi),\\
\lambda_{\perp,+}(p)&=\Lambda(\Phi)-[D(p)-|C(p)|].\nonumber
\end{align}
Here $E_0(p)$ denotes the same spiral energy as Eq.~\eqref{eq:Eopt}, evaluated at wave vector $p$. The upper transverse eigenvalue is not independent of the lower one: $\lambda_{\perp,+}(p)=\lambda_{\perp,-}(p)+2|C(p)|$.

Equation~\eqref{eq:perp_eigs} relates transverse stability directly to the exchange spectrum. A global exchange-energy minimum has a nonnegative lower transverse branch. The in-plane fluctuation energies provide the complementary stability test.

\subsection{Effect of easy-plane anisotropy}
An easy-plane anisotropy is included as
\begin{equation}\label{eq:anisotropy}
\mathcal H_K=K\sum_{n,\alpha=A,B}(e_{\alpha n}^{z})^2,\qquad K>0.
\end{equation}
Easy-plane anisotropy penalizes a spin component perpendicular to the spiral plane. It therefore increases the restoring energy for out-of-plane deviations without changing the in-plane result. In the normalization of Eq.~\eqref{eq:quadratic_in}, the transverse matrix gains $2K\mathsf I$. For a planar spiral $i$, all small out-of-plane rotations have nonnegative energy when
\begin{equation}\label{eq:Kcrit}
K\geq K_{c,i}=\frac12\left[E_0(\Phi_i)-\min_{p\in[-\pi,\pi]}E_0(p)\right].
\end{equation}
The critical anisotropy is nonnegative, and it vanishes for the global ground state, for which $E_0(\Phi_i)=\min_pE_0(p)$; Eq.~\eqref{eq:Kcrit} is therefore relevant for metastable spirals. At the threshold, one out-of-plane distortion still costs no energy to quadratic order. A positive restoring energy requires $K>K_{c,i}$. Small in-plane rotations are stable when the eigenvalues in Eq.~\eqref{eq:planar_eigs} are nonnegative throughout the Brillouin zone. This condition tests stability against small spin rotations; it does not determine the energy barriers between distinct magnetic states.

The harmonic spectra above refer to rotations about zero-field stationary spirals. An applied field changes the stationary spin configuration, so its stability must be examined using the Hessian evaluated at that field-dependent configuration. In particular, local stability of a planar solution is distinct from its equilibrium selection among competing planar and nonplanar states.

\section{Response to a magnetic field}\label{sec:field}
\subsection{Unrestricted isotropic ground state}
The field response depends on whether the spiral plane is free to rotate. We first consider the equilibrium response of the isotropic Hamiltonian, without constraining the spins to their zero-field plane. We denote the largest exchange eigenvalue by $\Lambda=\max_p\lambda_+(p)$ and its ordering wave vector by $Q$. For the exchange parameters considered here, $C(0)>0$, so the uniform ferromagnetic mode has eigenvalue $\lambda_+(0)=D(0)+|C(0)|$. A field $h$ along $z$ couples to both layers through $-h\sum_{n,\alpha}e^z_{\alpha n}$.

A conical spiral combines the optimal transverse exchange mode at $Q$ with a uniform magnetization $m$ along the field:
\begin{align}\label{eq:cone}
\bm e_{An}&=\bigl(s\cos nQ,s\sin nQ,m\bigr),\nonumber\\
\bm e_{Bn}&=\bigl(s\cos(nQ+\delta_Q),\nonumber\\
&\hspace{2.7em}s\sin(nQ+\delta_Q),m\bigr),
\end{align}
where $s=\sqrt{1-m^2}$ and $\delta_Q=-\arg C(Q)$. Its energy per two-layer repeat is
\begin{equation}\label{eq:coneenergy}
E_{\rm cone}(m,h)=-\Lambda+[\Lambda-\lambda_+(0)]m^2-2hm.
\end{equation}
The conical spiral is the classical magnetization process of an isotropic helimagnet~\cite{KitanoNagamiya1964}, and it is an exact ground state under the conditions stated here, for fields along the cone axis. At fixed uniform magnetization $m$, the Fourier-space exchange bound puts the uniform magnetization in the $p=0$ ferromagnetic eigenmode and bounds every remaining Fourier component by $\Lambda$. The conical spiral in Eq.~\eqref{eq:cone} has exactly this lowest exchange energy while preserving the length of each spin. Thus the cone is a global classical ground state for the present isotropic equal-diagonal exchange matrix when the uniform $p=0$ mode is ferromagnetic ($C(0)>0$). For other exchange parameters, the uniform mode may have a different sublattice character, and the conical result then requires a separate examination.

Minimizing Eq.~\eqref{eq:coneenergy} yields
\begin{align}\label{eq:hsat}
h_{\rm sat}&=\Lambda-\lambda_+(0),\\
m(h)&=\min\bigl(1,h/h_{\rm sat}\bigr),\qquad h\geq0,\nonumber\\
E_{\rm gs}(h)&=\begin{cases}-\Lambda-h^2/h_{\rm sat},&0\leq h\leq h_{\rm sat},\\-\lambda_+(0)-2h,&h\geq h_{\rm sat}.\end{cases}\nonumber
\end{align}
For a nonferromagnetic ground state with $h_{\rm sat}>0$, the equilibrium wave vector $Q$ is independent of field below saturation. Rotational invariance makes this result independent of the direction of the laboratory field. The magnetization per repeat is $2m$, so the corresponding susceptibility is $2/h_{\rm sat}$ below saturation. If the exchange ground state is already ferromagnetic, $\Lambda=\lambda_+(0)$ and $h_{\rm sat}=0$: the zero-temperature isotropic ferromagnet is fully polarized for any $h>0$, so the finite-saturation-field expressions $h/h_{\rm sat}$ and $2/h_{\rm sat}$ do not apply. A positive easy-plane anisotropy gives a separate, finite hard-axis saturation scale, as discussed below.

\subsection{Constrained in-plane response}
When the spiral plane is fixed and contains the field, individual layer angles can relax instead of forming the unrestricted cone. For a noncollinear stationary spiral, the harmonic Hessian in Eq.~\eqref{eq:Hparallel} determines its initial response. The magnetic field induces spin rotations with Fourier components at $p=\pm\Phi_i$; for a generic incommensurate pitch, the source vector at $p=\Phi_i$ is $\bm v_i=(1,e^{i\delta_i})^T$. The susceptibility follows from the quadratic change in exchange energy \emph{per two-layer repeat},
\begin{align}\label{eq:chi_exact}
E_i^{\rm planar}(h)&=E_0(\Phi_i)-\tfrac12\chi_i^{\rm planar}h^2+O(h^4),\\
\chi_i^{\rm planar}&=\tfrac12\bm v_i^\dagger\mathsf H_\parallel(\Phi_i)^{+}\bm v_i.\nonumber
\end{align}
Here $+$ denotes the inverse on the field-coupled subspace. A zero-energy mode coupled to the field produces a divergent harmonic response; a pseudoinverse does not remove that physical divergence. The factor $1/2$ follows from the two conjugate Fourier sources and the energy normalization per repeat; it is not the susceptibility per spin. Commensurate cases where the two source wave vectors coincide require separate treatment. The susceptibility obtained in this way does not establish whether the distorted planar state is stable against all spin rotations.

\subsection{Perpendicular field with easy-plane anisotropy}\label{sec:rmfieldcomparison}
RM define their magnetic energy per pair of layers. Their Zeeman contribution is $-BM_s(\cos\theta_A+\cos\theta_B)$, and their easy-plane anisotropy is $K(\cos^2\theta_A+\cos^2\theta_B)$, with $M_s$ and $K$ defined for one layer~\cite{RosenfeldMushnikov2008}. Our unit-spin field variable is therefore $h=BM_s$; in particular, the reduced field $BM_s/J_1$ used in RM's Fig.~4 is exactly $h/J_1$, without an additional factor of two.

Adding the easy-plane term in Eq.~\eqref{eq:anisotropy} to the isotropic cone of Eq.~\eqref{eq:coneenergy} gives the perpendicular-field response. The spins cant uniformly while retaining their zero-field pitch and relative layer angle. Writing $m=M/M_s=\cos\theta$, the energy per two-layer repeat is
\begin{equation}\label{eq:rmperp}
E_\perp(m,B)=-\Lambda+\bigl[\Lambda-\lambda_+(0)+2K\bigr]m^2-2BM_s m,
\end{equation}
where $\Lambda=\max_k\lambda_+(k)$. Minimization gives
\begin{align}\label{eq:rmhardaxis}
K_{\rm eff}&=K+\tfrac12[\Lambda-\lambda_+(0)],\\
B_{\rm sat}&=\frac{2K_{\rm eff}}{M_s},\qquad
m(B)=\min\left(1,\frac{BM_s}{2K_{\rm eff}}\right).\nonumber
\end{align}
Equations~\eqref{eq:rmperp} and \eqref{eq:rmhardaxis} reduce directly to RM's Eqs.~(8) and (9) when $J_4=J_5=J_6=0$. The additional exchanges change the effective anisotropy and saturation field, not the form of the canting response. The uniform cone is the exact classical ground state also in the presence of the easy-plane term. At fixed $m$ the cone attains the exchange bound, and $\sum_{n,\alpha}(e^z_{\alpha n})^2\geq(\sum_{n,\alpha}e^z_{\alpha n})^2/2N$, with equality only for uniform canting. The result assumes rigid equal-length layer moments, a common easy plane, and a uniform ferromagnetic $k=0$ mode.

For the three-exchange and six-exchange endpoints of Table~\ref{tab:progression}, Eq.~\eqref{eq:rmhardaxis} gives $K_{\rm eff}/J_1=0.0504$ and $0.0641$, respectively. The corresponding saturation fields, already listed in that table, differ by about $27\%$. The increase reflects the change in the exchange spectrum.

\subsection{In-plane field: six-exchange energy and limits}\label{sec:rmplanarfield}
For an in-plane field, let $a_n$ and $b_n$ be the independent planar angles of the two layers in repeat $n$. The direct six-exchange generalization of RM's Eq.~(10) is
\begin{widetext}
\begin{align}\label{eq:rmplanarsix}
E_\parallel=-\frac1N\sum_{n=0}^{N-1}\Bigl\{&BM_s[\cos a_n+\cos b_n]
+J_1\cos(b_n-a_n)+J_2\cos(a_{n+1}-b_n)\nonumber\\
&+J_3[\cos(a_{n+1}-a_n)+\cos(b_{n+1}-b_n)]\nonumber\\
&+J_4\cos(b_{n+1}-a_n)+J_5\cos(a_{n+2}-b_n)\nonumber\\
&+J_6[\cos(a_{n+2}-a_n)+\cos(b_{n+2}-b_n)]\Bigr\}.
\end{align}
\end{widetext}
The sum is over two-layer repeats with periodic or suitably commensurate boundary conditions. Our expression reduces to RM's planar energy for $J_4=J_5=J_6=0$. Here the spins are restricted to the plane, so their possible rotation out of that plane is excluded. A positive easy-plane anisotropy makes such configurations physically relevant, but only a sufficiently strong anisotropy stabilizes them against out-of-plane spin rotations. A minimum within the plane need not be stable in three dimensions.

\subsection{Recovery of the RM in-plane-field comparison}\label{sec:rmreproduction}
\begin{figure*}[t]
\centering
\includegraphics[width=0.90\textwidth]{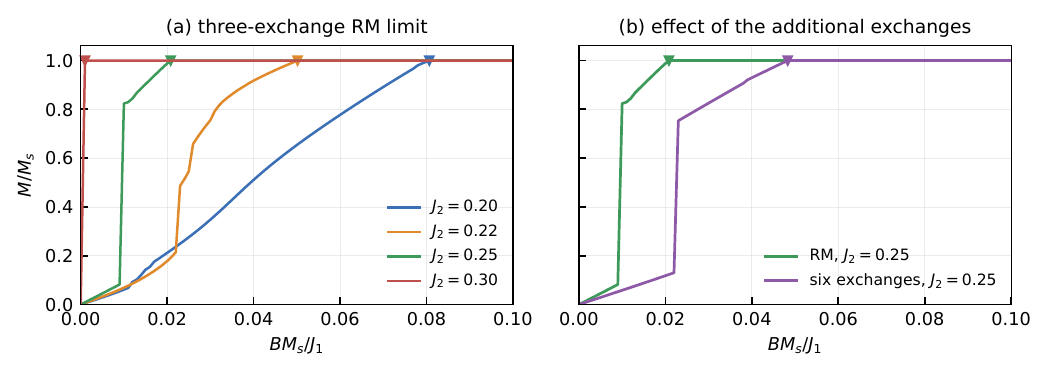}
\caption{Constrained in-plane magnetization. (a) RM limit for $J_1=1$, $J_3=-0.12$, and four values of $J_2$.
(b) Comparison at $J_2=0.25$ with the six-exchange extension $(J_4,J_5,J_6)=(0.025,0.045,-0.035)$. Each point is the
lowest energy found among six initial configurations on periodic chains selected to approximate the zero-field pitch (the allowed values are $q=k/N$). Triangles mark
$h=\Lambda-\lambda_+(0)$, where the in-plane ferromagnet becomes locally stable. The planar curves are sampled at intervals of $0.001J_1$; triangles indicate local stability rather than an equilibrium transition.}
\label{fig:rminplane}
\end{figure*}

Instead of assuming that the extended model reproduces the RM field response, we calculated the constrained-planar response of RM for $J_1=1$, $J_3=-0.12$, and $J_2=0.20,0.22,0.25,0.30$ with the six-exchange extension at $J_2=0.25$ and $(J_4,J_5,J_6)=(0.025,0.045,-0.035)$. For each parameter set, we minimized Eq.~\eqref{eq:rmplanarsix} on periodic chains chosen to approximate the infinite-chain pitch closely: $N=101$, 67, 70 and 104 for $J_2=0.20$, 0.22, 0.25 and 0.30, and $N=59$ for the six-exchange model. The residual mismatch between the optimal pitch and an allowed $k/N$ is below $2\times10^{-4}$; exact commensurability is not implied. A fixed chain length pins the zero-field pitch to $q=k/N$; for $N=40$ this changes the low-field slope by up to 7\% (Supplemental Material~\cite{SM}). At each of 101 fields in $0\leq h/J_1\leq0.10$, the lowest energy was taken from relaxations started from the three best commensurate spirals, the ferromagnet, a random configuration and the solution at the preceding field, using analytical gradients. The RM curves recover the qualitative low-field distortion and differing approaches to saturation reported in Ref.~\cite{RosenfeldMushnikov2008}. The triangles in Fig.~\ref{fig:rminplane} mark $h=\Lambda-\lambda_+(0)$, the local-stability threshold of the in-plane ferromagnet. This threshold need not coincide with the equilibrium transition field. The six-exchange example changes the preferred zero-field pitch from $q\simeq0.24281$ to $q\simeq0.20327$ and raises this field from $0.0208J_1$ to $0.0483J_1$. These results demonstrate the sensitivity of constrained-planar field response to longer-range exchange, but do not isolate individual contributions of $J_4,J_5$, and $J_6$. The plotted energy is the smallest found among the numerical starting configurations at each field, not a proof of the absolute planar minimum. Apparent jumps can reflect crossings of competing planar branches or finite-chain commensuration; assigning an equilibrium first-order transition requires converged branch energies, a system-size study, and stability against nonplanar rotations.

\paragraph{Transverse stability of field-distorted planar states.}
The in-plane relaxation fixes a stationary set of angles $\{\phi_{\alpha n}(h)\}$, but does not establish stability against rotations out of the easy plane. Let $\mathsf H_\perp(h,0)$ be the transverse Hessian of the exchange and Zeeman energy, evaluated at that stationary planar configuration with $K=0$, using the same quadratic-energy normalization as Eq.~\eqref{eq:quadratic_in}. Since the easy-plane anisotropy vanishes for every planar spin configuration, the stationary angles are independent of $K$, and the transverse Hessian at finite anisotropy is $\mathsf H_\perp(h,K)=\mathsf H_\perp(h,0)+2K\mathsf I$. Consequently, the minimum anisotropy required for \emph{local transverse nonnegativity} of this particular branch is
\begin{equation}\label{eq:Kcfield}
K_c(h)=\max\!\left\{0,-\tfrac12\lambda_{\min}[\mathsf H_\perp(h,0)]\right\}.
\end{equation}
This field-dependent condition extends Eq.~\eqref{eq:Kcrit} to nonuniform planar configurations. It is not an equilibrium $B$--$K$ phase boundary: in-plane stability and comparison with other planar and nonplanar states remain necessary. This transverse criterion is separate from the planar energy comparison and from the exact hard-axis conical response in Eqs.~\eqref{eq:rmperp} and \eqref{eq:rmhardaxis}.

\paragraph{Field response along the exchange sequence.}
Adding the exchange interactions in sequence also changes the spin-wave stability spectra and hard-axis magnetization. In particular, the five-exchange ferromagnet has $\Lambda=\lambda_+(0)$ and hence $h_{\rm sat}=2K$, whereas the sixth exchange restores a spiral and a finite exchange contribution to saturation. The full spectral and magnetization comparison is provided in the Supplemental Material~\cite{SM}; these hard-axis curves follow the exact conical solution and should not be confused with nonlinear in-plane field relaxation.

\section{Information obtainable from magnetic measurements}\label{sec:inverse}
A measured spiral pitch $q$, internal phase $\delta$, hard-axis saturation scale $h_{\rm sat}$, and constrained-planar susceptibility $\chi_{\parallel}$ provide four quantities for six exchange constants. At most four combinations of the exchanges can therefore be determined, and at least two remain free. For the representative six-exchange parameters, the $4\times6$ Jacobian of these observables has the maximal rank four, so exactly four combinations are constrained to first order. The sensitivity differs strongly between these combinations. With exchanges expressed in the units and parameterization used in the Supplemental Material, the singular values of the relative Jacobian $\partial\ln O_\mu/\partial J_a$ are 72.9, 9.3, 6.0 and 1.0. Their magnitudes and condition number depend on this choice of exchange coordinates and observable normalization; they do not by themselves quantify experimental uncertainties. The weakest combination changes all four observables by only about 1\% for a change of $0.01J_1$, so it is poorly determined by data of realistic precision. The Jacobian and its assumptions are given in the Supplemental Material~\cite{SM}. Wave-vector-resolved measurements, such as inelastic neutron scattering, or independent first-principles exchange information are therefore necessary.

\section{Discussion}\label{sec:discussion}
The influence of an interlayer exchange on the ordering wave vector depends on its contribution to the exchange spectrum, not simply on its distance or magnitude. Even relatively small additional couplings can change the preferred stacking. Consequently, collinear energy comparisons alone may miss the magnetic ground state of a layered kagome magnet.

Different measurements probe different aspects of this competition. The hard-axis saturation field measures the separation between the ordering and uniform exchange eigenvalues, together with the anisotropy contribution, whereas the in-plane response probes the relaxation of individual layer directions. The latter can involve competing stationary configurations, so the finite-chain curves cannot alone establish equilibrium field-induced transitions. The weakly constrained exchange combination found in Sec.~\ref{sec:inverse} further motivates momentum-resolved excitation measurements or independent first-principles exchange calculations.

The conclusions apply to the classical rigid-layer model. For specific $AT_6X_6$ compounds~\cite{Chen2026,ChenArxiv2026}, the validity of intralayer rigidity and the possible importance of other magnetic sublattices, anisotropic interactions, and longitudinal fluctuations must be assessed separately.

\section{Conclusions}\label{sec:conclusions}
For two equivalent magnetic layers per repeat, the Luttinger--Tisza construction gives an exact classical single-$Q$ ground state. In the six-exchange model, additional interlayer couplings can change the preferred order from spiral to ferromagnetic and back to spiral. The finite-chain planar field response illustrates competing configurations but does not, by itself, establish equilibrium first-order transitions.

The four magnetic observables considered here constrain at most four combinations of the six exchange constants, with one combination weakly determined. Determining the magnetic stacking therefore requires a sufficiently complete exchange description and comparison with noncollinear states; momentum-resolved measurements can supply further constraints.

\begin{acknowledgments}
Work at Ames National Laboratory was supported by the U.S. Department of Energy (DOE), Office of Science, Basic Energy Sciences, Materials Sciences and Engineering Division. Ames National Laboratory is operated for the U.S. DOE by Iowa State University under Contract No. DE-AC02-07CH11358.
\end{acknowledgments}

\end{document}